# Modelling proportionate growth

**Tridib Sadhu[§] and Deepak Dhar[¶]**

[§]Department of Physics of Complex Systems
Weizmann Institute of Science
Rehovot 76100
Israel.

[¶]Department of Theoretical Physics
Tata Institute of Fundamental Research
Mumbai 400005
India.

Key words: Proportionate growth; Biological growth; Pattern formation; Sandpile Model.

**Abstract:** *An important question in biology is how the relative size of different organs is kept nearly constant during growth of an animal. This property, called proportionate growth, has received increased attention in recent years. We discuss our recent work on a simple model where this feature comes out quite naturally from local rules, without fine tuning any parameter. The patterns produced are composed of large distinguishable structures with sharp boundaries, all of which grow at the same rate, keeping their overall shapes unchanged.*

It is fascinating to see baby animals grow into adults. Understanding the development of different organs from a single egg cell has been the central problem in developmental biology for over a hundred years. However, there is a considerably simpler problem of understanding how a small baby animal grows to a much larger size. In the case of humans, the body weight increases by a factor of 30 or so. In the case of elephants, this factor is about 100. As the baby grows, different parts of the body grow at same rate. This is called proportionate growth. Of course, this is only a good first approximation. For example, in humans, it is well known that the head grows less than the limbs, some changes in body structure occur at puberty, etc. However, at the simplest level of description, it is useful to ignore such complications.

Understanding how different organs are formed, starting from a single cell is the subject of cell differentiation and morphogenesis. The basic mechanism underlying this is believed to be the Turing instability in reaction-diffusion systems [1]. In a baby becoming an adult, all the organs are already formed, and we sidestep this more difficult question. We would like to emphasize that even the simpler question is not well-understood. The important point is that proportionate growth requires regulation, and coordination between the different growing parts. If there is no regulation, it would be very difficult to maintain the overall left-right symmetry that is seen in many animals. In a cell, all chemical reaction rates have a fair amount of fluctuation, because the number of molecules of the chemical species undergoing change is typically small. If the growth in different parts were independent, these fluctuations would lead to much larger variations in net growth than what is observed. For example, in mammals, the bilateral symmetry is maintained quite well during growth (typically to within a few percent).

That proportionate growth is special is clear from the fact that examples of proportionate growth outside the biological world are difficult to find. Sure, if one takes a balloon, with some picture drawn on it, and blows it up, all parts of the picture grow proportionately. But this is not really `growth', it is just stretching. Or, consider the growth in a droplet of water suspended in air supersaturated with water vapor. As the droplet collects more water from the surrounding air, it grows in size, and keeps its roughly spherical shape. But in this case, there are no internal distinct parts, and hence this also does not qualify as proportionate growth. One can think of crystals growing from a supersaturated solution. The crystals can have nontrivial shapes, but all the growth occurs on the surface, and the structure of the internal regions, once formed remains frozen. There are many other examples of growth studied in physics literature so far, e.g. diffusion limited aggregation [2], surface growth by molecular beam epitaxy [3], Eden growth model [4], invasion percolation [5] etc. In all these cases also, the structure of inner parts gets frozen, and growth occurs only at the surface, and not everywhere.

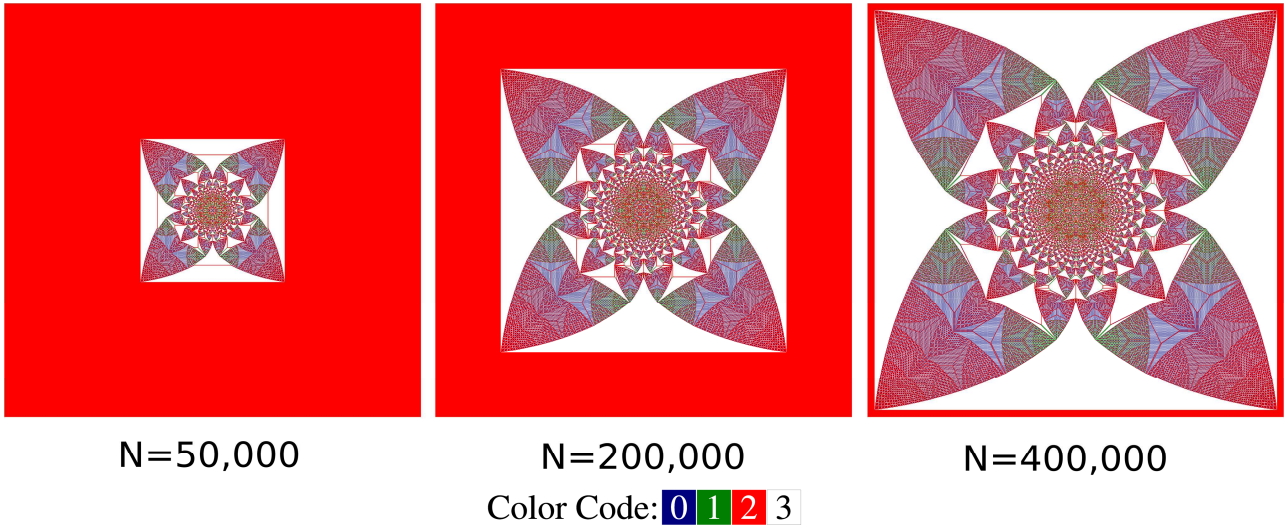

Figure 1: The pattern produced by adding N particles at a single site on a square lattice, and relaxing. Initial configuration is with all sites occupied by two particles. All three patterns are on a $1000 \times 1000$ lattice.

A biologist may object that the general mechanism of proportionate growth is quite well understood. The explanation typically only identifies some chemical agents, variously called hormones, promoters, inhibitors growth factors, that regulate the production of different proteins, and thus control growth. These agents themselves are produced or degraded by complicated regulatory processes that turn on or off different genes. This whole process is orchestrated by the genetic program encoded in the animal's DNA.

The role of DNA as chief controller of all processes in biological systems cannot be denied. But the DNA is a complicated molecule, as it has to do many more things than just ensuring proportionate growth. If we want only the latter, perhaps we can construct a simpler system that meets our objective, *which does not invoke the full complexity of DNA*. We discuss below one such model where the formation of complex structures and their proportionate growth is achieved with only a very small number of states per cell and the same set of instructions.

The model is called the abelian sandpile model, and it was introduced in 1987 by Bak, Tang and Wiesenfeld as a model of self-organized criticality [6]. This model has an interesting mathematical structure that makes it analytically tractable and it has been studied a lot (see [7], for a review). However, much of this work is related to studying the `criticality' of the model, reflected in the fact that it shows burst-like relaxation, with a power-law distribution of the size of avalanches. We want to emphasize the self-organization aspect here, which is a much older concept, and formed the bedrock for Bak et al's theory. In fact, the idea of self-organization itself came from attempts to understand biological growth. Thus, applying ideas of sandpiles back to biological growth, the theory has come a full circle.

We define the sandpile model on a two dimensional square lattice as follows: at each site of the lattice, there is a non-negative integer called the number of grains at that site. If the number of grains at a site exceeds 3, the site is said to be unstable, and it 'topples' by transferring four grains from that site, one to each neighbor. We start with an initial configuration where all sites are stable, and the grains form a periodic structure in space. For example, all sites could be having precisely two grains at the beginning.

At each time step, we add a grain to the site at the origin. If this makes the site unstable, we relax the configuration by toppling at that site. If this results in some other site becoming unstable, they are also relaxed, until all sites become stable. This concludes the time step. The name 'abelian' refers to the fact that we get the same final configuration irrespective of the order in which different unstable sites are relaxed. After the stable configuration is reached, a new grain is added at the origin, and so on.

Our model takes into account the basic phenomenology that the cell-division process operates under some threshold conditions: it does not happen until adequate resources are available. The addition of grains is like providing food, which is

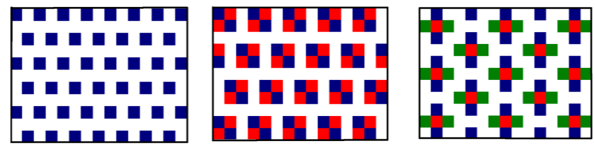

Figure 2: Examples of periodic distribution of grains inside patches in the pattern in Figure 1. Each colored square represent a site. The color code is same as used in the full pattern.

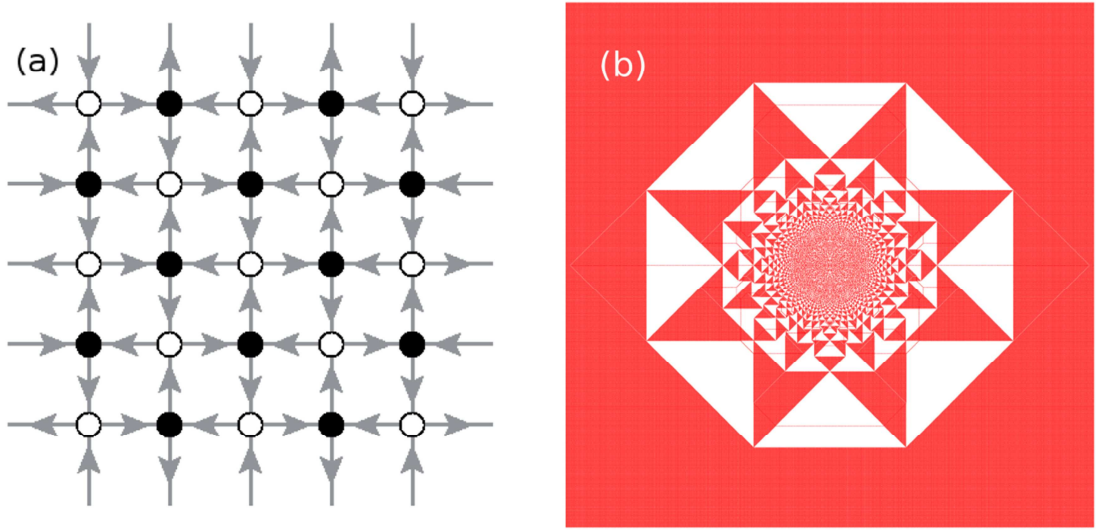

**Figure 3: (a) Checkerboard distribution of particles on the F-lattice. The filled circles denote occupied sites and the unfilled empty. (b) The pattern produced on this background. The color code: red=0, and white=1. The background pattern is a checkerboard pattern of red and white squares. The apparent uniform red color is an artifact of the reduced size of the figure. (Reproduced from [9])**

required for growth.

The Figure 1 shows the result of this evolution after N particles have been added. The three patterns correspond to N=50,000, 200,000, and 400,000, respectively. Each cell is given a color depending on the number of particles in the cell. The color code is shown in the figure. We see that the added particles get distributed over a square region, with some local rearrangement of grains. This region forms a colored pattern over the red background. This pattern becomes bigger as N is increased, and appears to show proportionate growth.

It is not necessary to work with the square lattice, or a particular initial background. One can take other lattices like the honeycomb or the triangular lattice. Even lattices in three or higher dimensions show similar result. Also, by changing the initial periodic background, we can get any number of different patterns. See [8, 9] for more such examples. The variety of aesthetically pleasing pictures that one can generate this way provided the main motivation behind our study of these patterns

There are other studies of patterns in sandpiles, which we mention briefly here. The first of such was [10], drawing attention to the intricate patterns. The periodic pattern inside patches was discussed in [11]. For more recent studies of growing sandpiles, and related models see [12, 13]. The connection to integrable models in statistical physics is explored in [14].

These pictures look like the pictures of fractals, but in the simplest cases on a two dimensional lattice the box dimension [3] of the disturbed region is 2. For some more complicated patterns, like in [9], this can be different and need not be integer. One can also see some repeated motifs in the pattern, for example inside the petal-like structures in Figure 1. The repeated motifs are also seen in fractals, but unlike them, these do not show self-similarity under scale transformations which is the defining characteristic of fractals.

For the pattern in Figure 1, it is easy to see that the length of the side of the square toppled region would grow as $\sqrt{N}$: we have added N extra particles, and as in a stable configuration, each square can accommodate at most 3 particles, the area of the region disturbed by added grains must grow at least linearly with N.

A closer look at the pattern shows many interesting features. The pattern is made up of a large number of patches; within each patch the arrangement of grains is periodic. The relative position, shape and fractional area of the patch appear to remain constant as N increases. In Figure 2, we have shown details of the periodic arrangements in some selected patches. The whole pattern then looks like a quilt made with pieces of cloth with different periodic patterns, sown together to form the full design. Alternatively, we may like to think of different patches as different 'organs' of the 'animal'.

While the periodic arrangement of grains within a patch is the most striking feature of the patterns, it is interesting that we do not have a rigorous proof, starting from the evolution rules of the model, of this property yet. For the present, we will take it as an observed property in the patterns studied so far.

In fact, the pattern in Figure 1 is rather complicated, and it is difficult to analyze. The pattern shown in Figure 3 is much simpler. It is produced on a lattice that is a variant of the square lattice called the F-lattice, shown in Figure 3a. The bonds of the square lattice have direction attached to them, and each site has two arrows coming into it, and two going out. The stable heights are only 0 and 1, and whenever the height exceeds 1, two grains

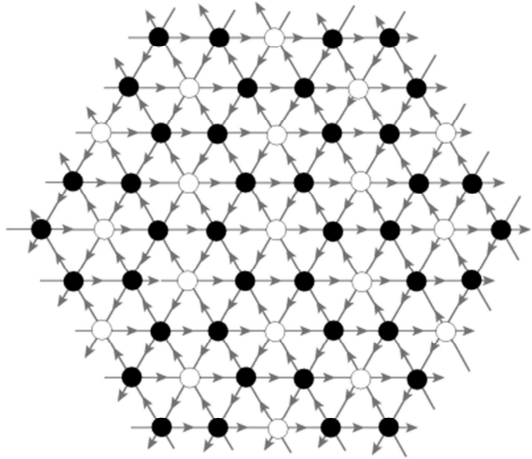 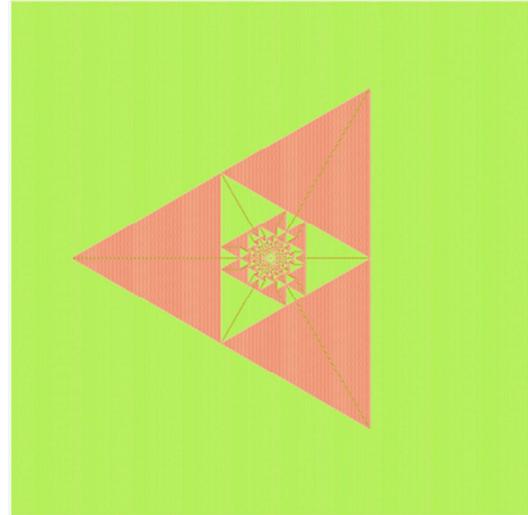

**Figure 4: (a) A periodic arrangement of particles on a triangular lattice with directed edges. The filled circles represent sites with one particle, and unfilled with two particles. (b) The pattern produced by adding N particles at a single site on the background in (a). (Reproduced from [9])**

are thrown out in the direction of outgoing arrows from that site. We start with an initial background where the occupied sites are arranged as the black squares on a checkerboard. The pattern shown in Figure 3b is produced by adding N=300,000 grains at a single site on this background, and relaxing.

This pattern is simpler, as there are only two stable heights at a site: 0 and 1. We find that there are only two types of patches in the pattern, those where all sites are occupied, and those with alternate sites occupied (as in the background pattern). All patch boundaries are straight lines, with all patches being 3- or 4-sided polygons. Also, the angles of the patch boundaries with the x-axis are only integer multiples of $\pi/4$. These facts and the observed adjacency structure of patches (i.e., which patches share a boundary) allowed us to obtain a detailed characterization of the asymptotic pattern for large N [8]. In particular, we find that the boundary of the asymptotic pattern is a regular octagon. This is an extra symmetry that emerges only in the large N limit.

We have studied the patterns produced on different backgrounds also. An important question is how the diameter Λ of the pattern depends on N. If we start with empty background, or when the mean density of particles in the background is low enough, on a d-dimensional lattice we get compact growth of the pattern with $\Lambda \sim N^{1/d}$. On the other hand, if the average density in a background is too high, one can get infinite avalanches, with Λ becoming infinite for finite $N$. This has been termed explosive growth [15].

Remarkably, we found an infinite class of non-explosive backgrounds, for which the diameter grows as $N^\alpha$ with $1/d \leq \alpha \leq 1$. An example of such a background on a triangular lattice with directed edges is shown in Figure 4a. The pattern produced on this background, after adding $N = 4000$ grains is shown in Figure 4b. For this pattern the growth exponent $\alpha = 1$. It turns out that this pattern is even easier to characterize than the pattern in Figure 3. In Figure 5, we have shown another fast-growing pattern, on F-lattice for which $\alpha \approx 0.725$. The details are skipped here, and may be found in [9].

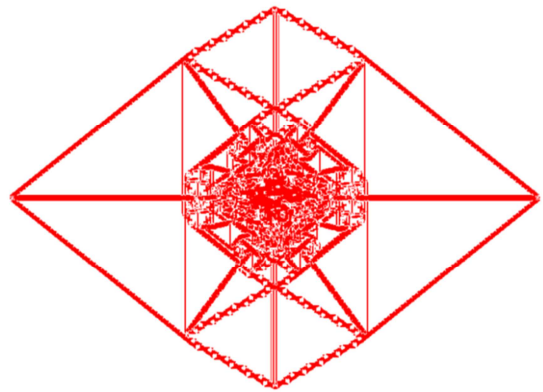

**Figure 5: The pattern produced on the F-lattice with a background formed by empty sites along the boundary of tilted rectangles, and rest filled. Only the boundaries of the resolved patches are shown here. (Reproduced from [9])**

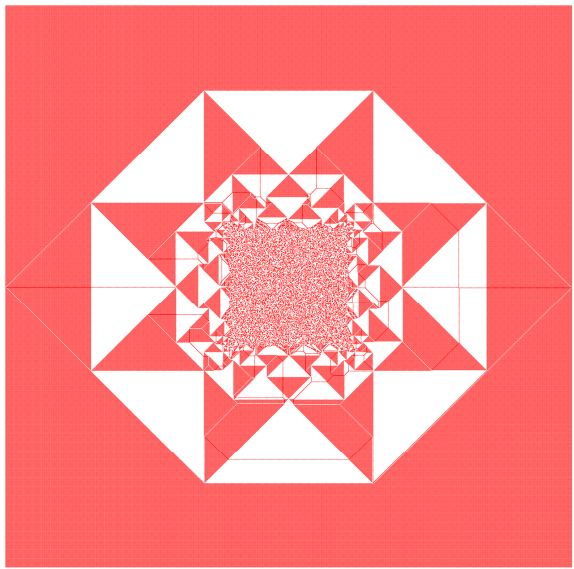

**Figure 6: The pattern produced on the F-lattice by adding N=100,000 particles at sites randomly chosen from a square region of width $b = 100$ lattice units. The initial configuration has a checkerboard distribution of heights 0 and 1. Color codes same as in Figure 3. (Reproduced from [16])**

An important feature of these sandpile patterns is their analytical tractability. We briefly indicate how these patterns are characterized quantitatively. We define the toppling function $T_N(x, y)$ which gives the number of topplings at site $(x, y)$, when $N$ particles are added at the origin, and the configuration is relaxed. We introduce the reduced coordinates $\xi = x/\Lambda$, and $\eta = y/\Lambda$. Then for large $N$, $T_N(x, y)$ has the scaling behavior $T \approx \Lambda^\beta g(\xi, \eta)$, where the scaling function $g$ specifies the asymptotic pattern. Let the diameter $\Lambda$ grows as $N^\alpha$. The fact that the function T has to be an integer function imposes very strong constraints on the function $g(\xi, \eta)$. It turns out that for compact patterns with $\alpha = 1/d$, we have $\beta = 2$, and the function $g$ is exactly quadratic in each patch. For $\alpha > 1/d$, we can only have $\beta = 1$, and the function $g$ has to be a linear function within a patch [9]. Then the continuity properties of $g$, and the known adjacency structure of the patches in the pattern gives a set of coupled linear equations for the coefficients of the polynomial. These can be solved exactly to determine the function $g$ completely and hence characterize the pattern.

One important issue is whether such cellular automaton models, like the sandpile model, with deterministic evolution rules can be considered as good models of biological growth, where noise has an important role. During growth the animal body is influenced by fluctuations of environmental and of local origin. Despite such fluctuations, the proportionate growth is maintained. It is thus important to study the robustness of these patterns to introduction of noise.

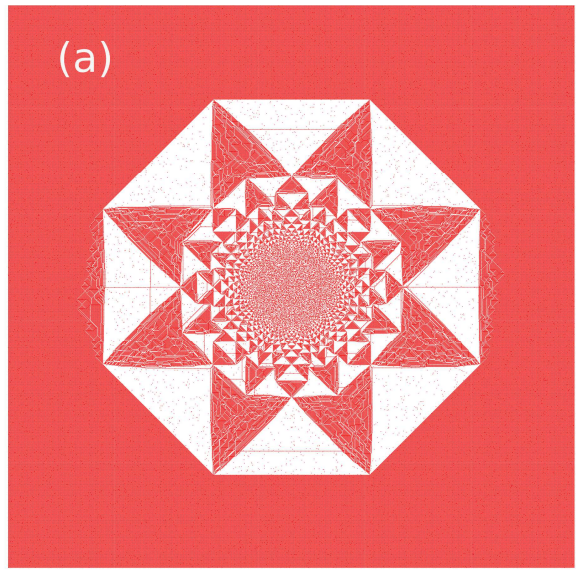

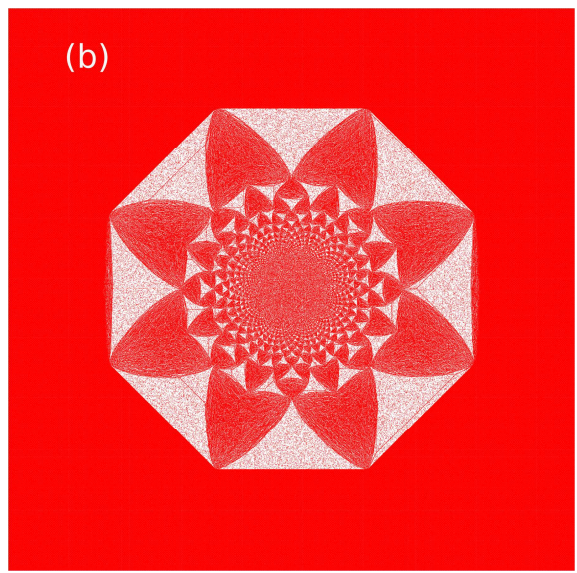

**Figure 7: The patterns produced on the F-lattice by adding large number of particles at a single site on a background of mostly checkerboard distribution, except for height 1 at (a) 1% and (b) 10% of the sites, respectively, being replaced by height 0. Color code: red =0, white=1. (Reproduced from [16])**

We have studied the effect of randomness in the point of addition on the patterns produced [16]. It turns out that a small amount of this randomness does not change the large-scale properties of the pattern. If the particles are added at a site chosen at each time step randomly from a small square of size $b$, then for $N \gg b^2$, the large scale structure of the pattern is unchanged. Only in a small area near the origin of width of order $b$, one can see the effect of the randomness, but the fractional size of this region decreases to zero as N increases to infinity. An example of such a pattern is shown in Figure 6.

We also studied the effect of noise in the initial pattern. If the background pattern is not exactly periodic, and a small fraction of sites start with incorrect initial number of grains, we find that for small noise, the system still shows proportionate growth, and the large-scale pattern is qualitatively unchanged. However, the exact ratios of sizes of different patches in the asymptotic pattern depend on the strength of the noise. Also the boundaries between patches are no longer straight-lines. In Figure 7, we have shown the effect of 1% and 10% noise in the initial configuration in the F-lattice pattern in Figure 3.

The third type of noise we studied was in the evolution rules. In this case, we assumed that at each toppling, there is a small probability that some particles are lost. It seems like even a small amount of noise of this type can destroy the pattern, and result in a circular featureless blob [16]. In other cases, some features of pattern persist for a long time, but other more intricate features are quickly smeared out, and lost [17].

Another interesting question is the effect of boundaries, or presence of dead sites on the lattice. We found that in presence of boundaries, the pattern is modified, but survives. Importantly, the rate at which diameter changes with N is also modified [18]. For example, for a compact pattern growing in half space, where any particle falling on the boundary is lost, the diameter increases not as $\sqrt{N}$, but as $N^{1/3}$.

While many examples of simple evolution rules giving rise to complex structures are known (the game of Life [19] is a well-known example), in most cases these patterns are not tractable analytically. The patterns we described in this article are special in that they are of intermediate complexity, and are analytically tractable. In fact, the exact characterization of the asymptotic pattern involves some interesting mathematics: discrete analytic functions, and piece-wise linear functions. In this article, we have not discussed these. The interested reader is referred to [9] for a discussion.

There are many interesting open problems. The most obvious is the already mentioned absence of a proof of the proportionate growth property. Also, we cannot predict, at present, which periodic structures within patches are possible for a given background pattern. The observed stability of patterns in presence of noise also needs explanation, as the scaled toppling function is no longer piece-wise linear, or piece-wise quadratic. For the fast growing patterns, the analytical determination of the fractal dimension $\alpha$ is a challenging problem.

TS acknowledge the support of Israel Science foundation. DD would like to acknowledge partial financial support from the Department of Science and Technology, Government of India, through a JC Bose Fellowship.

# 1. Bibliography

[1] M. C. Cross and P. C. Hohenberg, "Pattern formation outside of equilibrium," *Rev. Mod. Phys.,* vol. 65, p. 851, 1993.

[2] T. A. Witten and L. M. Sander, "Diffusion-limited aggregation, a kinetic critical phenomenon," *Phys. Rev. Lett.,* vol. 47, p. 1400, 1981.

[3] H. E. Stanley and A. L. Barabási, Fractal Concepts in Surface Growth, Cambridge University Press, 1995.

[4] M. Eden, "A two-dimensional growth process," *Proc. 4th Berkeley Symp. Math. Stat. Probab.,* vol. 4, p. 223, 1961.

[5] D. Wilkinson and J. F. Willemsen, "Invasion percolation: a new form of percolation theory," *J. Phys. A: Math. Gen.,* vol. 16, p. 3365, 1983.

[6] P. Bak, C. Tang and K. Wiesenfeld, "Self-organized criticality: An explanation of the 1/f noise," *Phys. Rev. Lett.,* vol. 59, p. 381, 1987.

[7] D. Dhar, "Theoretical studies of self-organized criticality," *Physica A,* vol. 369, no. 1, p. 29, 2006.

[8] D. Dhar, T. Sadhu and S. Chandra, "Pattern formation in growing sandpiles," *Europhys. Lett.,* vol. 85, p. 48002, 2009.

[9] T. Sadhu and D. Dhar, "Pattern formation in fast-growing sandpiles," *Phys. Rev. E,* vol. 85, p. 021107, 2012.

[10] S. H. Liu, T. Kaplan and L. J. Gray, "Geometry and dynamics of deterministic sandpiles," *Phys. Rev. A,* vol. 42, p. 3207, 1990.

[11] S. Ostojic, "Patterns formed by addition of grains to only one site of an abelian sandpile," *Physica A,* vol. 318, p. 187, 2003.

[12] A. Fey and F. Redig, "Limiting shapes for deterministic centrally seeded growth models," *J. Stat. Phys.,* vol. 130, p. 579, 2008.

[13] L. Levine and Y. Peres, "Strong spherical asymptotics for rotor-router aggregation and the divisible sandpile," *Potential Analysis,* vol. 30, p. 1, 2009.

[14] S. Caracciolo, G. Paoletti and A. Sportiello, "Conservation laws for strings in the Abelian Sandpile Model," *Europhys. Lett.,* vol. 90, p. 60003, 2010.

[15] A. Fey, L. Levine and Y. Peres, "Growth Rates and Explosions in Sandpiles," *J. Stat. Phys.,* vol. 138, p. 143, 2010.

[16] T. Sadhu and D. Dhar, "The effect of noise on patterns formed by growing sandpiles," *J. Stat. Mech.,* p. P03001, 2011.

[17] R. S. Dandekar and D. Dhar, *Unpublished.*

[18] T. Sadhu and D. Dhar, "Pattern Formation in Growing Sandpiles with Multiple Sources or Sinks," *J. Stat. Phys.,* vol. 138, p. 815, 2010.

[19] L. S. Schulman and P. E. Seiden, "Statistical mechanics of a dynamical system based on Conway's game of Life," *J. Stat. Phys.,* vol. 19, p. 293, 1978.